\documentclass[pra,amsmath,amssymb]{revtex4}

\begin{document}

\title{Programs in Mathematica relevant to Phase Integral Approximation for 
coupled  ODEs of the Schr{\"o}dinger type}

\author{A. A. Skorupski}
\email[]{askor@fuw.edu.pl}
\affiliation{Department of Theoretical Physics, Soltan Institute for Nuclear
Studies, Ho\.za 69, 00--681 Warsaw, Poland
            }


\begin{abstract}
Three programs in Mathematica are presented, which produce expressions for the
lowest order and the higher order corrections of the Phase Integral
Approximation. First program is pertinent to one ordinary differential equation
of the Schr{\"o}dinger type. The remaining two refer to a set of two such
equations.
\end{abstract}

\maketitle

\section{\label{intr}Introduction}

In this paper we present three programs in Mathematica related to the Phase
Integral Approximation (PIA) \cite{genpia}. They produce expressions for the
lowest order aproximation and the higher order corrections. The first program
generates the higher order corrections $Y_{2n}(x)$ pertinent to one ordinary
differential equation of the Schr{\"o}dinger type. The second program gives the
vectors $\mathbf{b}_m(x)$ and the third one the corrections $Y_m(x)$ and the
vectors $\mathbf{s}_m(x)$. These quantities are pertinent to a set of two 
ordinary differential equation of the Schr{\"o}dinger type. The programs will be
referred to as \verb%Y2n, bvm% and  \verb%Ymsvm%, respectively. They should
be saved in \verb%file.mat% (\verb%file% equal to \verb%Y2n, bvm% or \verb%Ymsvm%)
and run by using the standard Mathematica input command \verb%<< file.mat%.
The user supplied input data should be saved in \verb%file.dat%. If
\verb%file.dat% is an empty file, the default input data contained in each
program will be used in computation.

Each program opens a dialog in Mathematica session in which the user
is asked to specify the form of output from Mathematica (\verb%OutputForm%, 
\verb%TeXForm% or \verb%FortranForm%), give the number of terms to be determined
and answer a few questions specific to each program.  The results produced
by Mathematica will be saved as \verb%file.res%, \verb%file.resTeX% or
\verb%file.resFor%. All equation and section numbers given in what follows
refer to \cite{genpia}.

One should answer \verb%y% to the dialog question
\verb%Expand[ , Trig -> True ],  y/n?%, if the user supplied data file contains
some trig functions, and \verb%n% otherwise.

The factor $g(x)$ after
the dialog question \verb%Factor g[x] in eigenvector = ?% in the program
\verb%Ymsvm%, can be chosen either to be equal to one, or to the denominator
in the just printed expression for \verb%s02/s01% (= $s_{02}(x)/s_{01}(x)$),
see Eq.~(124). In the last case, the implied factors should also be included,
e.g., $\sin x$ comming from  $\tan x$ etc.

All three programs deal with multiple sums, see Eqs.~(40) and (54).
These sums are programmed in the simplest possible way, e.g., the sum
\[
\sum_{\substack{
	\alpha+\beta+\gamma+\delta+\sigma=m\\
	\sigma \geq 1}} Y_{\alpha} Y_{\beta} Y_{\gamma} Y_{\delta} \,
\mathbf{s}_{\sigma}
\]
present in Eq.~(54), where $0 \leq \alpha, \beta, \gamma, \delta, \sigma \leq
m-1$, is programmed as

\noindent
\begin{verbatim}
  sum2 = 0;
  Do[ sum2 += If[ a + b + g + d + s == m, (*then*)
    Y[a] Y[b] Y[g] Y[d] sv[s], (*else*) 0 ],
    {a, 0, m - 1}, {b, 0, m - 1}, {g, 0, m - 1}, {d, 0, m - 1}, {s, 1, m - 1} ];
\end{verbatim}
etc. This makes the programs as close as possible to mathematical formulas
thereby eliminating programming errors. For the same reasons, the
integral which defines the coordinate $(\mathbf{e},\mathbf{s}_m)$, see
Eq.~(105), was not simplified by using Eqs.~(106)--(108) which would make the
computation faster. However, this would make programming a bit more complicated
and  error-prone. In our computations, this type od optimalizations was not
necessary. Our aim was to produce correct results in a reasonable time (seconds
or minutes rather than hours). A strong test for the correctness of programming
was the vanishing of the odd order corrections, $Y_{2n-1}(x) \equiv 0$ (which
required cancellation of many terms), see Sec.~VIIIA. Another check was the fact
that in hermitian cases, the same results were produced in the simplified
hermitian and non-hermitian theory, see Secs.~VIIIA--VIIIC and VIIIE.

\newpage

\section{Program to determine the corections $Y_{2n}$ from the recurrence
relation (40)}

\noindent
File \verb%Y2n.mat%:
\begin{verbatim}
(*******************************************************************************

  Calculation of the phase integral corrections Y2n from recurrence relations,
  Eq. (40) in [1], for a scalar case, i.e. for one ODE of the Schroedinger type:

                         u''(x) + R(x) u(x) = 0.

  [1]  A. A. Skorupski, "Phase Integral Approximation for coupled ODEs of the
       Schroedinger type", arXiv: 0710.5868, Sec. II.

********************************************************************************

*****************  Define type of output from Mathematica  *********************
*)
outpform = InputString["Output, TeX or Fortran form of results,  o/t/f? "];
sc = If[ outpform == "o", OpenWrite["Y2n.res", FormatType -> OutputForm],
If[ outpform == "t", OpenWrite["Y2n.resTeX", FormatType -> TeXForm],
OpenWrite["Y2n.resFor", FormatType -> FortranForm ] ] ];
(**)
outpYm   = InputString[
            "Simple fractions or Common denominator or in Y2n,  s/c? "];
(**)
WriteString[sc, "\n Formulas for corrections Y2n as functions of x or z
(= zeta variable). \n"];
(*
******************   Define maximum value of n in Y[2 n]   *********************
*)
nmax = Input["nmax = ? "];
WriteString[sc, "\n nmax = "]; Write[ sc, nmax ];
(*
*****************   Define type of input to Mathematica    *********************
*)
inptform = InputString["Input of R(x) and a(x) or General Y2n,  i/g? "];
(**)
t0 = TimeUsed[];
(**)
If [ inptform == "i",
(*then*)
  (*
  **********************  Define default input data  ***************************
  *)
  (***   Parabolic Model   ***)
  af = 0;
  R  = coef (x^2 - x1^2);
  (*
  ****************  Read new input data from file Y2n.dat  *********************
  *)
  << Y2n.dat;
  WriteString[sc, "\n Y2n[x] for \n"];
  (*
  *************************  Write input data  *********************************
  *)
  WriteString[sc, "\n R[x] = \n"]; Write[sc, R];
  WriteString[sc, "\n Auxiliary function a[x] = \n"]; Write[sc,  af];
  (**)
  (****************************************************************************)
  Qsq = R - af; dQsq = D[ Qsq, x ];
  Qsqor1 = Qsq;
  ep0 = ( (5/16) (dQsq/Qsq)^2 - (1/4) D[ dQsq, x ]/Qsq + af )/Qsq;
  aux  = Simplify[ ep0 ];
  aux1 = Together[aux];
  WriteString[sc, "\n eps0[x] = \n"]; Write[sc, aux1],
(*else*)
  (**********    Prepare quantities for general calculations     **************)
  (**)
  ep0 = eps0[x]; Qsq = Qsqr[x];
  xorz = InputString["x or zeta variable,  x/z? "];
  If[ xorz == "x", (*then*) WriteString[sc,
  "\n Y2n[x] as functions of eps0[x], Qsqr[x] = Q^2[x] and derivatives \n"];
  Qsqor1 = Qsq, (*else*)
  Qsqor1 = 1; x = z; WriteString[sc,
  "\n Y2n[z] as functions of eps0[z] and derivatives \n"]; ]
  ];
(**)
Qm2 = 1/Qsqor1;
Y[0] = 1;
(*
**********************   Start iterations for Y[2 n]   *************************
*)
For[n = 1, n <= nmax, n++,
  sum1 = 0; sum2 = 0; sum3 = 0;
  m = 2 n;
  Do[ sum1 += If[ a + b == m, (*then*) Y[a] Y[b], (*else*) 0 ],
    {a, 0, m - 2, 2}, {b, 0, m - 2, 2} ];
  Do[ sum2 += If[ a + b + g + d == m, (*then*)
    Y[a] Y[b] Y[g] Y[d], (*else*) 0 ],
    {a, 0, m - 2, 2}, {b, 0, m - 2, 2}, {g, 0, m - 2, 2}, {d, 0, m - 2, 2} ];
  Do[ sum3 += If[ a + b == m - 2, (*then*)
    ep0 Y[a] Y[b] + (3/4) Qm2 D[Y[a], x] D[Y[b], x] -
    (1/2) Y[a] Qm2 (D[Y[b], {x, 2}] - (1/2) Qm2 D[ Qsqor1, x] D[Y[b], x] ),
    (*else*) 0 ], {a, 0, m - 2, 2}, {b, 0, m - 2, 2} ];
(**)
  Y[m] = (1/2) (sum1 - sum2 + sum3);
(**)
   ];
t = TimeUsed[];
WriteString[sc, "\n CPU time used for computation (seconds) = "];
Write[sc, t - t0];
(*
***********************  Simplify and write results  ***************************
*)
For[n = 1, n <= nmax, n++,
    m = 2 n;
    WriteString[sc, "\n n = "]; Write[sc, n];
    aux  = Simplify[ Y[m] ];
    aux1 = If[ outpYm == "c", Together[aux], Apart[aux, x] ];
    WriteString[sc, "\n Y2n = \n"]; Write[ sc, aux1 ]
   ];
t = TimeUsed[];
WriteString[sc,
"\n CPU time used for computation & simplification (seconds) = "];
Write[sc, t - t0];
\end{verbatim}
End of file \verb%Y2n.mat%.

The file that follows is an example of the data file for the program
\verb%Y2n%. As it stands it is an ampty file containing only commemts. By
uncommenting the definition of the functions $a(x)$ and $R(x)$:
\verb%(*% $\to$ \verb%(**)% and \verb%*)% $\to$ \verb%(**)%, one activates
the input data.\\[1ex]

\noindent
File \verb%Y2n.dat%:
\begin{verbatim}
(* Data pertinent to R[x] in the differential equation u''[x] + R[x] u[x] = 0.
   By default the auxiliary function af[x] = 0. For other choice
   include the data command: af = your_function[x];
*) 
(************************   Budden's Model:   *********************************)
(*
af = 0;
R  = coef x/(x - p);
*)
(******************************************************************************)
\end{verbatim}
End of file \verb%Y2n.dat%.

\section{Program to determine $\mathbf{b}_m$ from the recurrence relation (54)}

\noindent
File \verb%bvm.mat%:
\begin{verbatim}
(*******************************************************************************

       Calculation of bv[m] from the recurrence relation, Eq. (54) in [1]			 
             
[1]  A. A. Skorupski, "Phase Integral Approximation for coupled ODEs of the
     Schroedinger type", arXiv: 0710.5868, Sec. III.

********************************************************************************
*)
bvm = OpenWrite["bvm.res", FormatType -> OutputForm];
(**)
WriteString[bvm, "\n Formulas for vectors bvm as functions of x or z
(= zeta variable). \n"];
t0 = TimeUsed[];
Unprotect[Sqrt, Power];
Sqrt[x_^2] := x;
i^n_ := (-1)^(n/2) /; EvenQ[n];
i^n_ := i (-1)^((n-1)/2) /; OddQ[n];
re[x_] := Coefficient[x, i, 0];
im[x_] := Coefficient[x, i, 1];
cc[x_] := re[x] - i im[x];
Potect [Sqrt, Power];
(*
********************   Define maximum value of m in bv[m]   ********************
*)
mmax = Input["\n mmax = ? "];
(**)
xorz = InputString["\n x or zeta variable,  x/z? "];
If[ xorz == "z", (*then*) Q[x_] := 1; x = z ];
Qm1 = Q[x]^(-1); Qm2 = Qm1^2; dQ = D[ Q[x], x ];
(**)
Y[x, 0] = 1;
bv[1] = i Qm1 D[sv[x, 0], x];
Y1eq0Q = InputString["\n Y[x, 1] = 0,  y/n? "];
If[ Y1eq0Q == "y", (*then*) Y[x, 1] = 0 ];
(*
************************   Start iterations for bv[m]   ************************
*)
For[m = 2, m <= mmax + 1, m++,
(**)
  sum1 = 0; sum2 = 0; sum3 = 0; sum4 = 0; sum5 = 0; sum6 = 0;
  Do[ sum1 += If[ a + b + s == m, (*then*)
    Y[x, a] Y[x, b] ( sv[x, s] + 2 (Y[x, s] sv[x, 0] - bv[s]) ), (*else*)
    0 ], {a, 0, m - 1}, {b, 0, m - 1}, {s, 1, m - 1} ];
  Do[ sum2 += If[ a + b + g + d + s == m, (*then*)
    Y[x, a] Y[x, b] Y[x, g] Y[x, d] sv[x, s], (*else*) 0 ],
    {a, 0, m - 1}, {b, 0, m - 1}, {g, 0, m - 1}, {d, 0, m - 1}, {s, 1, m - 1} ];
  Do[ sum3 += If[ a + b == m, (*then*) Y[x, a] Y[x, b], (*else*) 0 ],
    {a, 1, m - 1}, {b, 1, m - 1} ];
  Do[ sum4 += If[ a + b + g + d == m, (*then*)
    Y[x, a] Y[x, b] Y[x, g] Y[x, d], (*else*) 0 ],
	{a, 0, m - 1}, {b, 0, m - 1}, {g, 0, m - 1}, {d, 0, m - 1} ];
  Do[ sum5 += If[ a + b + g + s == m - 1, (*then*)
    Y[x, a] Y[x, b] Y[x, g] Qm1 D[sv[x, s], x], (*else*) 0 ],
	{a, 0, m - 1}, {b, 0, m - 1}, {g, 0, m - 1}, {s, 0, m - 1} ];
  Do[ sum6 += If[ a + b + s == m - 2, (*then*)
    Y[x, a] ( Y[x, b] ( Qm2 ( D[sv[x, s], {x, 2}] - Qm1 dQ D[sv[x, s], x ] ) +
	eps0[x] sv[x, s] ) - Qm2 D[Y[x, b], x] D[sv[x, s], x] -
    (1/2) Qm2 ( D[Y[x, b], {x, 2}] - Qm1 dQ D[Y[x, b], x ] ) sv[x, s] ) +
    (3/4) Qm2 D[Y[x, a], x] D[Y[x, b], x] sv[x, s], (*else*) 0 ],
	{a, 0, m - 2}, {b, 0, m - 2}, {s, 0, m - 2} ];
(**)
  bv[m] = (1/2) (sum1 - sum2 + ( sum3 - sum4 ) sv[x, 0] + i 2 sum5 + sum6);
(**)
   ];
t = TimeUsed[];
WriteString[bvm, "\n CPU time used (seconds) = "]; Write[bvm, t - t0];
(**)
(*
***********************  Simplify and write results  ***************************
*)
For[m = 1, m <= mmax, m++,
  WriteString[bvm, "\n m = "]; Write[bvm, m];
  aux  = Simplify[ bv[m] ];
  aux1 = Together[aux];
  WriteString[bvm, "\n bvm = \n"]; Write[ bvm, aux1 ];
(**)
   ];
\end{verbatim}
End of file \verb%bvm.mat%.

\section{Program to determine $Y_m$ and $\mathbf{s}_m$ from recurrence relations
for 2 ODEs with either hermitian or non-hermitian matrix, see [1], Sec. VI}

\noindent
File \verb%Ymsvm.mat%:
\begin{verbatim}
(*******************************************************************************

           Calculation of Y[m] and sv[m] from recurrence relations
         for 2 ODEs with either hermitian or non-hermitian matrix [1].

[1]  A. A. Skorupski, "Phase Integral Approximation for coupled ODEs of the
     Schroedinger type", arXiv: 0710.5868, Sec. VI.

********************************************************************************
*)
outpform = InputString["Output, TeX or Fortran form of results,  o/t/f? "];
vc = If[ outpform == "o", (*then*) OpenWrite["Ymsvm.res",
FormatType -> OutputForm],
(*else*) If[ outpform == "t", (*then*) OpenWrite["Ymsvm.resTeX",
FormatType -> TeXForm], (*else*) OpenWrite["Ymsvm.resFor",
FormatType -> FortranForm] ] ];
t0 = TimeUsed[];
Unprotect[Sqrt, Power];
Sqrt[x_^2] := x;
i^n_ := (-1)^(n/2) /; EvenQ[n];
i^n_ := i (-1)^((n-1)/2) /; OddQ[n];
re[x_] := Coefficient[x, i, 0];
im[x_] := Coefficient[x, i, 1];
cc[x_] := re[x] - i im[x];
Potect [Sqrt, Power];
(*
***************  Define maximum value of m in Y[m] and sv[m]  ******************
*)
mmax = Input["mmax = ? "];
WriteString[vc, "\n mmax = "]; Write[vc, mmax];
mmxp1 = mmax + 1;
(*
***********************  Define default input data  ****************************

One must define the auxiliary function a(x, p) -> af and the matrix elements
Rjk(x, p) -> Rjk,  where p represents parameter(s). By default, the variable
automatic = True. In that case, one must define a list of meaningful numerical
replacements: parrepls = { x -> x0, p -> p0 } which is necessary to calculate
automatically the variables: Delta given by Eq. (121), sqrtDel (= Sqrt[Delta])
and signQsq (= sign of Qsq given by Eq. (121)). If one puts automatic = False
in the data file Ymsvm.dat, the definitions of Delta, sqrtDel and signQsq must
be given in the file Ymsvm.dat.
*)
(*** Example A ***)
af  = 0;
R11 = x Cos[x]^2 + Sin[x]^2;
R12 = (x - 1) Cos[x] Sin[x];
R21 = (x - 1) Cos[x] Sin[x];
R22 = x Sin[x]^2 + Cos[x]^2;
parrepls = { x -> 2 };
automatic = True;
(*
*****************  Read new input data from file Ymsvm.dat  ********************
*)
<< Ymsvm.dat;
(*
**************************  Write input data  **********************************
*)
WriteString[vc, "\n R11 = \n"]; Write[vc, R11];
WriteString[vc, "\n R12 = \n"]; Write[vc, R12];
WriteString[vc, "\n R21 = \n"]; Write[vc, R21];
WriteString[vc, "\n R22 = \n"]; Write[vc, R22];
WriteString[vc, "\n  af = \n"]; Write[vc,  af];
If[ automatic, (*then*)
  WriteString[vc, "\n paramter replacement list = \n"]; Write[vc, parrepls],
  (*else*)
  WriteString[vc, "\n *** Non-automatic calculation *** \n"]
  ];
(**)
G11 = R11 - af; G12 = R12; G21 = R21; G22 = R22 - af;
(**)
exptrig = InputString["Expand[ , Trig -> True ],  y/n?  "];
QsqmQ   = InputString["Qsqr with minus or plus Sqrt[ Delta ],  m/p?  "];
WriteString[vc, "\n Qsqr with "];
If[ QsqmQ == "m", WriteString[vc, "minus "], WriteString[vc, "plus "] ];
WriteString[vc, "Sqrt[ Delta ] "];
(**)
(*
************  Find eigenvalues and eigenvectors of the G matrix  ***************
*)
If[ automatic, (*then*) Delta = (G11 - G22)^2 + 4 G12 G21 ];
If[ exptrig == "y", Delta = Expand[ Delta, Trig -> True ] ];
Delta = Factor[ Delta ];
WriteString[vc, "\n Delta = \n"]; Write[vc, Delta];
If[ automatic, (*then*) sqrtDel = Sqrt[ Delta ] ];
(**)
Qsq = (1/2) (G11 + G22 + If[ QsqmQ == "m", (*then*) -sqrtDel, (*else*)
sqrtDel ]);
Qsq = Simplify[ Qsq ];
If[ exptrig == "y", Qsq = Expand[ Qsq, Trig -> True ] ];
(**)
If[ automatic, (*then*)
  signQsq = If[ ( Qsq /. parrepls ) < 0, (*then*) -1, (*else*) 1,
  (*and if neither True or False then*) 1 ] ];
WriteString[vc, "\n signQsq = "]; Write[vc, signQsq];
WriteString[vc, "\n Qsq = \n"]; Write[vc, Qsq];
(**)
(*** eps0 = Simplify[ (Qsq^(1/4) D[Qsq^(-1/4), {x, 2}] + af)/Qsq ]; ***)
eps0 = Together[ Simplify[ ( (5/16) ( D[Qsq, x]/Qsq)^2 -
                        (1/4) D[Qsq, {x, 2}]/Qsq + af )/Qsq ] ];
WriteString[vc, "\n eps0 = \n"]; Write[vc, eps0];
(* << eps0.mat; *)
(**)
Q = If[ signQsq < 0, (*then*) - i (- Qsq)^(1/2), (*else*) Qsq^(1/2) ];
WriteString[vc, "\n Q = \n"]; Write[vc, Q];
(**)
Qm1 = Q^(-1); Qm2 = Qm1^2; dQ = D[ Q, x ];
(**)
s02os01 = ( Qsq - G11 )/G12;
If[ exptrig == "y", s02os01 = Expand[ s02os01, Trig -> True ] ];
s02os01 = Simplify[ s02os01 ];
WriteString[vc, "\n s02/s01 = \n"]; Write[vc, s02os01];
Print[ "s02/s01 = ", s02os01 ];
(**)
fact = Input["Factor g[x] in eigenvector = ? "];
WriteString[vc, "\n Factor g[x] = \n"]; Write[vc, fact];
s0v1 = fact; s0v2 = fact s02os01;
(**)
asqr = s0v1 cc[s0v1] + s0v2 cc[s0v2];
If[ exptrig == "y", asqr = Expand[ asqr, Trig -> True ] ];
(**)
normeigv = InputString["Normalized eigenvector,  y/n?  "];
If[ normeigv == "y", (*then*)
  ms0v = Sqrt[ Simplify[ asqr ] ];
  s0v1 = s0v1/ms0v; s0v2 =  s0v2/ms0v;
  asqr = 1;
  Print[ "\n(s0v(x), s0v'(x)) = " ];
  intg = cc[s0v1] D[ s0v1, x ] + cc[s0v2] D[ s0v2, x ];
  If[ exptrig == "y", intg = Expand[ intg, Trig -> True ] ];
  intg = Simplify[intg];
(**  Print["intg = ", intg];  **)
  If[ intg =!= 0, (*then*) (* Print[ "No, (s0v, s0v'(x)) =" ]; *) Print[ intg ];
    intheta = InputString[
    "Integrate (s0v(x), s0v'(x)) dx to calculate theta,  y/n?  "];
    If[ intheta == "y", (*then*)
      theta = i Integrate[ intg, x ];
      Print["theta = ", theta];
      WriteString[vc, "\n theta = \n"]; Write[vc, theta];
      phasf = Cos[theta] + i Sin[theta];
      s0v1 = s0v1 phasf; s0v2 = s0v2 phasf
      ], (*else*) Print[0]
    ];
  ];
s0v1 = Simplify[ s0v1 ]; s0v2 = Simplify[ s0v2 ];
s0v = { s0v1, s0v2 };
spv = { - cc[s0v2], cc[s0v1] };
(**)
WriteString[vc, "\n s0v = \n"]; Write[vc, s0v];
WriteString[vc, "\n spv = \n"]; Write[vc, spv];
WriteString[vc,
"\n ***  sv_m = cp_m spv + c_m s0v,  m = 1, 2, ..., mmax  *** \n"];
(**)
s0v1ms = s0v1 cc[s0v1]; s0v2ms = s0v2 cc[s0v2];
den = s0v1ms (G22 - Qsq) + s0v2ms (G11 - Qsq) -
      cc[s0v1] s0v2 G12 - s0v1 cc[s0v2] G21;
If[ exptrig == "y", den = Expand[ den, Trig -> True ] ];
den = Apart[ Simplify[ den ] ];
WriteString[vc, "\n D = \n"]; Write[vc, den];
(**)
coef = - 2 Qsq/den;
(**)
Y[0] = 1; sv[0] = s0v;
(**)
bv[1] = i Qm1 D[sv[0], x];
hrmthQ = InputString["Hermitian or Non-hermitian theory,  h/n?  "];
WriteString[vc,
If[ hrmthQ == "h", " ***  Hermitian ", " ***  Non-hermitian "] ];
WriteString[vc, "theory  *** \n"];
simthQ = If[ hrmthQ == "h", (*then*) InputString[
         "Simplified, Fulling or Wronskian conserving theory,  s/f/w?  "],
         (*else*) "s" ];
wresQ  = InputString["Write results,     y/n?  "];
sresQ  = InputString["Simplify results,  y/n?  "];
If[ wresQ == "y", (*then*)
outpYm = InputString[
"Simple fractions, Common denominator or NO output spec. in  Ym,  s/c/n?  "];
outpsv = InputString[
"Simple fractions, Common denominator or NO output spec. in svm,  s/c/n?  "] ];
aprog  = InputString["Append program,  y/n?  "];
(*
*******************  Start iterations for Y[m] and sv[m]  **********************
*)
For[m = 2, m <= mmxp1, m++,
  m1 = m - 1;
(**)
  cpf[m1] = coef { - s0v2, s0v1 } . bv[m1];
  If[ exptrig == "y", cpf[m1] = Expand[ cpf[m1], Trig -> True ] ];
  Print[ " m = ", m1 ];
  intgnt[m1] = If[ simthQ == "s", 0, If[ OddQ[m1] && simthQ == "w",
    2 re[ cpf[m1] { cc[D[s0v1, x]], cc[D[s0v2, x]] } . spv ],
  i 2 im[ cpf[m1] { cc[D[s0v1, x]], cc[D[s0v2, x]] } . spv ] ] ];                     
  If[ exptrig == "y", intgnt[m1] = Expand[ intgnt[m1], Trig -> True ] ];
(**)
  If[ m1 > 1, (*then*)
    For[ alpha = 1, alpha <= m1-1, alpha++,
         intgnt[m1] -= If[ simthQ == "s", 0, If[ OddQ[alpha] && simthQ == "w",
         -1, 1 ] {cc[sv[alpha][[1]]], cc[sv[alpha][[2]]]} . D[sv[m1 - alpha],
         x ] ]         
       ]
    ];
  If[ exptrig == "y", intgnt[m1] = Expand[ intgnt[m1], Trig -> True ] ];
  intgnt[m1] = Simplify[ intgnt[m1] ];
(*
  WriteString[vc, "\n m = \n"]; Write[vc, m1];
  WriteString[vc, "\n integrant[m] = \n"]; Write[vc, intgnt[m1]];
*)
(**)
  cf[m1] = If[ simthQ == "s", 0, Integrate[ intgnt[m1], x ] ];
  sv[m1] = cpf[m1] spv + cf[m1] s0v;
  If[ exptrig == "y", sv[m1] = Expand[ sv[m1], Trig -> True ] ];
(***  sv[m1] = Simplify[ sv[m1] ];  ***)
(**)
  Y[m1] = If[ hrmthQ == "h",
              (*then*) ( { cc[s0v1], cc[s0v2] } . bv[m1])/asqr,
              (*else*) (Qm2 cpf[m1] G12 asqr/(2 s0v1) +
              bv[m1][[1]])/s0v1 ];
  If[ exptrig == "y", Y[m1] = Expand[ Y[m1], Trig -> True ] ];
(*  WriteString[vc, "\n Y[m] = \n"]; Write[vc, Y[m1]];  *)
(***  Y[m1] = Simplify[ Y[m1] ];  ***)
(**)
  If[ m < mmxp1, (*then*)
  sum1 = 0; sum2 = 0; sum3 = 0; sum4 = 0; sum5 = 0; sum6 = 0;
  Do[ sum1 += If[ a + b + s == m, (*then*)
    Y[a] Y[b] ( sv[s] + 2 (Y[s] sv[0] - bv[s]) ), (*else*)
    0 ], {a, 0, m - 1}, {b, 0, m - 1}, {s, 1, m - 1} ];
  Do[ sum2 += If[ a + b + g + d + s == m, (*then*)
    Y[a] Y[b] Y[g] Y[d] sv[s], (*else*) 0 ],
    {a, 0, m - 1}, {b, 0, m - 1}, {g, 0, m - 1}, {d, 0, m - 1}, {s, 1, m - 1} ];
  Do[ sum3 += If[ a + b == m, (*then*) Y[a] Y[b], (*else*) 0 ],
    {a, 1, m - 1}, {b, 1, m - 1} ];
  Do[ sum4 += If[ a + b + g + d == m, (*then*)
    Y[a] Y[b] Y[g] Y[d], (*else*) 0 ],
    {a, 0, m - 1}, {b, 0, m - 1}, {g, 0, m - 1}, {d, 0, m - 1} ];
  Do[ sum5 += If[ a + b + g + s == m - 1, (*then*)
    Y[a] Y[b] Y[g] Qm1 D[sv[s], x], (*else*) 0 ],
    {a, 0, m - 1}, {b, 0, m - 1}, {g, 0, m - 1}, {s, 0, m - 1} ];
  Do[ sum6 += If[ a + b + s == m - 2, (*then*)
    Y[a] ( Y[b] ( Qm2 ( D[sv[s], {x, 2}] - Qm1 dQ D[sv[s], x ] ) +
    eps0 sv[s] ) - Qm2 D[Y[b], x] D[sv[s], x] -
    (1/2) Qm2 ( D[Y[b], {x, 2}] - Qm1 dQ D[Y[b], x ] ) sv[s] ) +
    (3/4) Qm2 D[Y[a], x] D[Y[b], x] sv[s], (*else*) 0 ],
    {a, 0, m - 2}, {b, 0, m - 2}, {s, 0, m - 2} ];
(**)
  bv[m] = (1/2) (sum1 - sum2 + ( sum3 - sum4 ) sv[0] + i 2 sum5 + sum6)]
(**)
   ];
t = TimeUsed[];
WriteString[vc, "\n CPU time used for computation (seconds) = "];
Write[vc, t - t0];
(**)
If[ wresQ == "y", (*then*)
(*
**********************  Simplify and/or write results  *************************
*)
For[m = 1, m <= mmax, m++,
  WriteString[vc, "\n m = "]; Write[vc, m];
  aux  = If[ sresQ == "y", (*then*) Simplify[ Y[m] ], (*else*) Y[m] ];
  aux1 = If[ outpYm == "c", Together[aux], If[ outpYm == "s", Apart[aux],
         aux ] ];
  WriteString[vc, "\n Y_m = \n"]; Write[ vc, aux1 ];
(**)
  aux  = cpf[m];
  aux1 = If[ exptrig == "y", Expand[aux, Trig -> True], aux ];
  aux2 = If[ sresQ == "y", (*then*) Simplify[aux1], (*else*) aux1 ];
  aux3 = If[ outpsv == "c", Together[aux2], If[ outpsv == "s", Apart[aux2],
         aux2 ] ];
  WriteString[vc, "\n cp_m = \n"]; Write[ vc, aux3 ];
(**)
  aux  = cf[m];
  aux1 = If[ exptrig == "y", Expand[aux, Trig -> True], aux ];
  aux2 = If[ sresQ == "y", (*then*) Simplify[aux1], (*else*) aux1 ];
  aux3 = If[ outpsv == "c", Together[aux2], If[ outpsv == "s", Apart[aux2],
         aux2 ] ];
  WriteString[vc, "\n c_m = \n"];  Write[ vc, aux3 ]
   ]
  ];
(**)
If[ aprog == "y", (*then*) << ap.mat ];
t = TimeUsed[];
WriteString[vc,
"\n CPU time used for computation & simplification (seconds) = "];
Write[vc, t - t0];
\end{verbatim}
End of file \verb%Ymsvm.mat%.\\[1ex]

The file that follows is an example of the data file for the program
\verb%Ymsvm%. Again it is an empty file containing only comments. By
uncommenting appropriate pieces, one can activate input data pertaining to
examples given in \cite{genpia}, Sec.~VIII.\\[1ex]

\noindent
File \verb%Ymsvm.dat%:
\begin{verbatim}
(**** Data for program Ymsvm ****)
(**)
(*** Example B ***)
(*
af = 0;
R11 = - ( x Cos[x]^2 + Sin[x]^2 );
R12 = (x - 1) Cos[x] Sin[x];
R21 = (x - 1) Cos[x] Sin[x];
R22 = - (x Sin[x]^2 + Cos[x]^2 );
parrepls = { x -> 2 };
*)
(*** Example C.1 ***) 
(*
af = 0;
R11 = x Cos[x]^2 + Sin[x]^2;
R12 =   i (x - 1) Cos[x] Sin[x];
R21 = - i (x - 1) Cos[x] Sin[x];
R22 = x Sin[x]^2 + Cos[x]^2;
parrepls = { x -> 2 };
*)
(*** Example C.2 ***)
(*
af = 0;
R11 = - ( x Cos[x]^2 + Sin[x]^2 );
R12 =   i (x - 1) Cos[x] Sin[x];
R21 = - i (x - 1) Cos[x] Sin[x];
R22 = - ( x Sin[x]^2 + Cos[x]^2 );
parrepls = { x -> 2 };
*)
(*** Example C.3 ***)
(*
af  = 0;
R11 = - ( x Cos[x]^2 + Sin[x]^2 );
R12 = (1 + i)/2^(1/2) (x - 1) Cos[x] Sin[x];
R21 = (1 - i)/2^(1/2) (x - 1) Cos[x] Sin[x];
R22 = - (x Sin[x]^2 + Cos[x]^2 );
parrepls = { x -> 2 };
*)
(*** Example C.4 ***)
(*
af  = 0;
R11 = - ( x Cos[x]^2 + Sin[x]^2 );
R12 = (Cos[fi] + i Sin[fi]) (x - 1) Cos[x] Sin[x];
R21 = (Cos[fi] - i Sin[fi]) (x - 1) Cos[x] Sin[x];
R22 = - (x Sin[x]^2 + Cos[x]^2 );
(*fi  = x;*)
parrepls = { x -> 2 };
*)
(*** Example D ***)
(*
af = 0;
R11 = x Cos[x]^2 + Sin[x]^2;
R12 = 2 i (x - 1) Cos[x] Sin[x];
R21 = - (1/2) i (x - 1) Cos[x] Sin[x];
R22 = x Sin[x]^2 + Cos[x]^2;
parrepls = { x -> 2 };
*)
(*** Example E ***)
(*
af = 0;
R11 = h0[x] - h1[x];
R12 = h2[x];
R21 = h2[x];
R22 = h0[x] + h1[x];
automatic = False;
r[x_] := Sqrt[ h1[x]^2 + h2[x]^2 ];
Delta = 4 r[x]^2;
sqrtDel = 2 r[x];
signQsq = - 1;
*)
(*** Example X ***)
(*
af = 0;
R11 = E1 - x^2;
R12 = x;
R21 = x;
R22 = E2 - 4 x^2;
parrepls = { x -> 3, E1 -> 1, E2 -> 2 };
*)
\end{verbatim}
End of file \verb%Ymsvm.dat%.\\[1ex]

The file that follows contains an example of a program that can be appended
to \verb%Ymsvm% by answering \verb%y% to the dialog question
\verb%Append program, y/n?%.
This program is pertinent to \verb%Example E% in the data file \verb%Ymsvm.dat%.
It computes and prints both the formulas in Eq.~(174) and all numerical results
given in TABLE I in \cite{genpia}. For each of two eigenvalues (for the minus or
plus sign in Eq.~(123)), the program
\verb%Ymsvm% should be run twice for two possible answers to the dialog question
\verb%Normalized eigenvector,  y/n?%. In both cases one should take 
\verb%mmax = 2% and answer \verb%n% to the dialog questions
\verb%Write results,  y/n?% and \verb%Simplify results,  y/n?%. Otherwise,
the program would spend very long time in an attempt to simplify $Y_2(x)$ and
$c_2^{\bot}(x)$ which in any case are too complicated to be presented.\\[1ex]

\noindent
File \verb%ap.mat%:
\begin{verbatim}
(***** Appended computation for program Ymsvm, Example E *****)
(**)
aux = Together[ Y[1] ];
WriteString[ vc, "\n Y_1 = \n"]; Write[ vc, aux ];
(**)
aux = Together[ cpf[1] ];
WriteString[vc, "\n cp_1 = \n"]; Write[ vc, aux ];
(**)
d0[x_] := 1/(4 x^2) + 4/x^4 + 38/x^6 + 748/x^8;
d1[x_] := 1/x^2 + 2/x^4 + 19/x^6 + 374/x^8;
h0[x_] := -1 - k^2 + d0[x]; h1[x_] := 2 (om + 1/x^2); h2[x_] := -1 + d1[x];
(**)
parrepls = { x -> 55, k -> 4/100, om -> 26041/10^7 };
WriteString[ vc, "\n"]; Write[ vc, parrepls ];
WriteString[ vc, "\n Q = "]; Write[ vc, N[ Q /. parrepls ] ];
WriteString[ vc, "\n eps0/2 = "]; Write[ vc, N[ eps0/2 /. parrepls ] ];
WriteString[ vc, "\n Y_1 = "]; Write[ vc, N[ Y[1] /. parrepls ] ];
WriteString[ vc, "\n Y_2 = "]; Write[ vc, N[ Y[2] /. parrepls ] ];
WriteString[ vc, "\n cp_1 = "]; Write[ vc, N[ cpf[1] /. parrepls ] ];
WriteString[ vc, "\n cp_2 = "]; Write[ vc, N[ cpf[2] /. parrepls ] ];
\end{verbatim}
End of file \verb%ap.mat%.\\[1ex]

\end{document}